\documentclass[prb,twocolumn,amsmath,amssymb,superscriptaddress,longbibliography]{revtex4-1}

\usepackage[normalem]{ulem}
\usepackage{graphicx}
\usepackage{todonotes} 
\usepackage{dcolumn}
\usepackage{tabularx}
\usepackage{bm}
\usepackage{hyperref}
\usepackage{textcomp}
\usepackage{verbatim}
\hypersetup{
    colorlinks=true,
    linkcolor=blue,
    citecolor = blue,
    filecolor=magenta,      
    urlcolor=blue,
    pdftitle={Thermoelectric response of fluxons in superconductors},
    pdfpagemode=FullScreen,
    }

\def\be{\begin{equation}}
\def\ee{\end{equation}}
\def\r{(\mathbf{r})}

\usepackage{multirow}
\usepackage{nicefrac}
\usepackage{qcircuit}
\usepackage{physics}

\usepackage{mathtools}

\begin{document}


\title{Giant Thermoelectric Response of Fluxons in Superconductors}

\author{Alok Nath Singh}
\affiliation{Department of Physics and Astronomy, University of Rochester, Rochester, NY 14627, USA}
\affiliation{Institute for Quantum Studies, Chapman University, Orange, CA 92866, USA}

\author{Bibek Bhandari}
\affiliation{Institute for Quantum Studies, Chapman University, Orange, CA 92866, USA}

\author{Alessandro Braggio}
\affiliation{NEST Istituto Nanoscienze-CNR and Scuola Normale Superiore, I-56127, Pisa, Italy}

\author{Francesco Giazotto}
\affiliation{NEST Istituto Nanoscienze-CNR and Scuola Normale Superiore, I-56127, Pisa, Italy}
\author{\\Andrew N. Jordan}
\affiliation{Institute for Quantum Studies, Chapman University, Orange, CA 92866, USA}
\affiliation{Department of Physics and Astronomy, University of Rochester, Rochester, NY 14627, USA}
\affiliation{The Kennedy Chair in Physics, Chapman University, Orange, CA 92866, USA}




\maketitle


\vspace{1em}

\noindent 
{\bf Thermoelectric devices that operate on quantum principles have been under extensive investigation in the past decades.\cite{Heremans2013when,Gehring2021complete,Brantut2013Thermoelectric,Zhang2020observation,Sothmann2015thermoelectric} 
These devices are at the fundamental limits of miniaturized heat engines and refrigerators, advancing the field of quantum thermodynamics\cite{Pekola2015toward,Whitney2014most,benenti2017fundamental,sothmann2012rect,williams2011effective,Jordan2013power,tesser2022heat,bibek2021minimal,sanchez2015chiral,sothmann2014quantum,scharf2020topological,scharf2021thermo,vischi2019thermodynamic,iorio2021photonic}. Most research in this area concerns the use of conduction electrons and holes as charge and heat carriers, and only very recently have superconductors been considered as thermal engines and thermoelectric devices\cite{connor2016resolving,Germanese2022bipolar,karmakar2024cyclic,Bergeret2018nonequilibrium,shimizu2019giant,manikandan2019super,hofer2016quantum,marchegiani2016self,marchegiani2020non,marchegiani2020super,ozaeta2014,germanese2023,de2023superconducting}.
 Here, we investigate the thermoelectric response of an Abrikosov vortex\cite{abrikosov1957magnetic} in type-II superconductors in the deep quantum limit. We consider two thermoelectric geometries, a type-II SIN junction and a local Scanning Tunneling Microscope (STM)-tip normal metal probe over the superconductor. We exploit the strong breaking of particle-hole symmetry in bound states at sub-gap energies within the superconducting vortex\cite{bardeen1969structure} to realize a giant thermoelectric response in the presence of fluxons. 
We predict a thermovoltage of a few mV/K at sub-Kelvin temperatures using both semi-analytic and numerical self-consistent solutions of the Bogoliubov-de Gennes equations. Relevant thermoelectric coefficients and figures of merit are found within our model, both in linear and nonlinear regimes. The ZT of the SIN junction is around 1, rising to above 3 for the STM junction centered at the vortex core.
We also discuss how this system can be used as a sensitive thermocouple, diode, or localized bolometer to detect low-energy single photons.
}

 BCS superconductors are usually seen as paradigmatic particle-hole symmetric systems where the linear thermoelectric response is not expected\cite{benenti2017fundamental} unless one considers a thermophase effect\cite{ginzburg1944thermoelectric,connor2016resolving}.  However, in type II superconductors, regions pierced by magnetic flux lines, called Abrikosov vortices, or fluxons, particle-hole symmetry may locally be broken\cite{bardeen1969structure}.  
The {\it sub-gap} excitations of the superconductor, corresponding to solutions of the Bogoliubov-de-Gennes (BdG) equations, are localized in the core of the vortex. Inside the fluxons, the Andreev reflection mechanism develops a superposition of particle-hole states, or vortex-bound states (VBS)\cite{caroli1964bound,de1999superconductivity}, around the vortex, characterizing the local low-energy behavior.
Indeed, the BdG equations self-consistently describe the winding of the superconducting order parameter around the magnetic flux line in the mean-field approximation.  Thus, the superconducting order parameter winding in the fluxon core determines a nontrivial topology to the electronic states.

In this article, we investigate the low-temperature quantum limit\cite{hayashi1998low} where the thermal energy is small compared to the level spacing of the VBS. This limit can be reached more easily for low electron density superconductors with low $k_F\xi_0$\cite{shan2011observation,Zhang2009nonmagnetic}. In these conditions, the asymmetry that emerges from the intrinsic doping of the material is reflected in particle or hole asymmetry of VBS, which in turn induces a strong thermoelectric response at low energies. The particle-hole symmetry breaking in the quantum limit has been experimentally observed in Ref.~\onlinecite{shan2011observation,nishimori2004}.

\begin{figure*}[t]
    \centering
    \includegraphics[width=\textwidth]{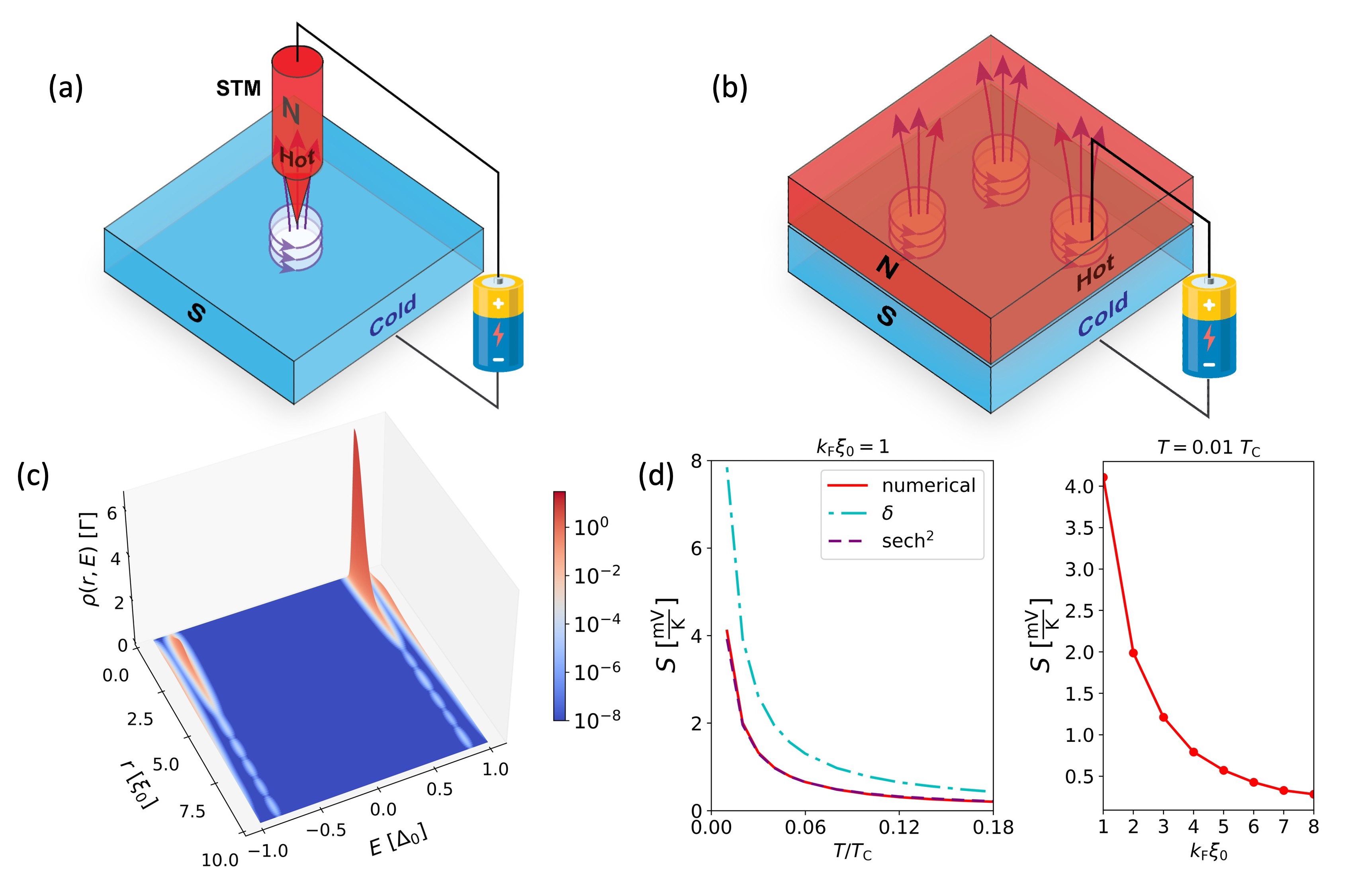}
    \caption{\textbf{(a)} A heated STM tip can be used to measure the localized thermoelectric response around the cold vortex core. \textbf{(b)} An SN junction to measure a broader thermoelectric current response involving a vortex lattice. \textbf{(c)} Local density of states plotted vs distance $r$ from the vortex core, and energy in the subgap region, in the units of $\Gamma = 10^2(\xi_0^2\Delta_0)^{-1}$. The dimensionless parameter $k_{\rm F}\xi_0$ equals $1$, and the normalized temperature $T/T_{\rm C}$ equals $0.01$. Particle/hole asymmetry is most pronounced at $r=0$, the vortex core. \textbf{(d)} Seebeck coefficient plotted versus normalized temperature (left plot) for $k_{\rm F}\xi_0 = 1$. The numerical result is plotted with the Dirac-$\delta$ and $\sech^2$ approximation for the local density of states peaks. On the right plot, the numerical result for the Seebeck coefficient is plotted against $k_{\rm F}\xi_0$, at a temperature of $0.01\ T_{\rm C}$.}
    \label{Fig1}
\end{figure*}

We propose two experimental setups to measure the thermoelectric effects, shown in Fig.~\ref{Fig1} (a) and (b). First is an STM-tip-like junction where the STM tip lies on top of the vortex core, giving a localized thermoelectric response that directly measures locally the contribution of a single vortex. Second is a Superconductor-Normal metal (SN) junction, separated by a thin insulating barrier that uses the whole extent of the vortex lattice (many vortices), giving higher conductivity and an increased power factor.

\vspace{1em}
\noindent\textbf{Bogoliubov-de-Gennes Equations}

\noindent We use the analysis done by de-Gennes \cite{de1999superconductivity} to describe the region near a vortex in type-II superconductors, at low temperatures and magnetic field strength just above the first critical value $H_{\rm C1}$ (isolated vortex). We express the Hamiltonian in terms of electron annihilation and creation operators in position basis ($\Psi(\mathbf{r}\alpha), \Psi^\dagger(\mathbf{r}\alpha)$) instead of the momentum basis, since translation symmetry is broken in the presence of a vortex. The mean field approximation is used to write the interaction terms in terms of one-particle interactions. Bogoliubov transformations are then used to diagonalize the Hamiltonian in terms of the quasiparticle operators, which gives us the Bogoliubov-de-Gennes (BdG) equations,
\be
\begin{aligned}
&(H_e + U(\mathbf{r}))u_n(\mathbf{r}) + \Delta(\mathbf{r})v_n(\mathbf{r}) = E_nu_n(\mathbf{r}),\\
-&(H_e + U(\mathbf{r}))v_n(\mathbf{r}) + \Delta^*(\mathbf{r})u_n(\mathbf{r}) = E_nv_n(\mathbf{r}).
\end{aligned}
\label{BdG}
\ee
$H_e = \frac{1}{2m}(-i\hbar\nabla - \frac{e\mathbf{A}}{c})^2 - E_F$ is the effective single-particle Hamiltonian, and $u_n\r,v_n\r$ are the quasiparticle amplitudes defined by the Bogoliubov transformation\cite{de1999superconductivity}. $U(\mathbf{r}) = -V\sum_n\left[|u_n(\mathbf{r})|^2f_n + |v_n(\mathbf{r})|^2(1-f_n)\right]$ and $\Delta(\mathbf{r}) = V\sum_nv_n(\mathbf{r})^*u_n(\mathbf{r})(1-2f_n)$ are the mean field and pairing potentials ($f_n$ is the Fermi function at energy $E_n$), which are dependent on the BCS attractive potential $V$. $U\r$ is nearly temperature-independent and can be approximated by the Hartree-Fock potential in the normal state. Thus, it can be incorporated into the chemical potential. However, $\Delta$ is nonzero around the Fermi surface and is strongly temperature dependent. Note that the mean-field Hamiltonian does not conserve the number of particles, so it needs to be done self consistently by keeping $N = 2\int d^3r\sum_n[|u_n\r|^2f_n + |v_n\r|^2(1-f_n)]$ constant.

\vspace{1em}
\noindent\textbf{Numerical and Analytical Solutions}

\noindent As motivated in the introduction, it is critical to break particle-hole symmetry to have a linear thermoelectric response.  
The BdG Hamiltonian generally possesses tenfold symmetry, including particle-hole, time-reversal, and chiral symmetry\cite{ranjith2019}. However, in the presence of a magnetic field, which results in a winding order parameter ($\Delta(\mathbf{r}) = |\Delta(\mathbf{r})|e^{i\theta}$, with $\theta$ being the azimuthal coordinate), time-reversal symmetry is broken. Moreover, the eigenstates of the BdG Hamiltonian with a winding order parameter have particle-hole asymmetric quasiparticle amplitudes ($u_n$, $v_n$), especially at the vortex core, where only the particle-like solution dominates.

In our case of interest - the single fluxon - the order parameter $\Delta$ has a phase winding number of 1 as a $2\pi$ rotation is made.  Thus, $u_n, v_n$ must then differ by one unit of angular momentum.  This leads to strong particle-hole symmetry breaking for low-lying excitations - either with\cite{kato2000quasi} or without\cite{hayashi1998low} the imposition of particle number conservation  - indicating a very robust phenomenon in the quantum limit. The quantum limit is given by
$k_{\rm B}T \leq \Delta_0^2/E_F \approx 3.528 k_{\rm B}T_c/k_{\rm F}\xi_0$, where $\Delta_0^2/E_{\rm F}$ gives the order of level spacing between the bound states\cite{caroli1964bound}, $k_{\rm F}$ is the Fermi wavenumber of the superconducting material, and $\xi_0$ is its coherence length. $T$ ($T_{\rm C}$) is the temperature (critical temperature) of the superconductor. Further, $\Delta_0$ is the superconducting gap at $T=0$. Beyond the quantum limit, the discrete nature of the bound states gets thermally smeared, leading to a substantial reduction of particle-hole symmetry breaking. From the quantum limit condition, it is evident that the asymmetry in the bound states manifests more clearly for materials with low values of $k_{\rm F}\xi_0$\cite{kato2000quasi,hayashi1998low} 
and high critical temperature $T_{\rm C}$.

\begin{figure*}[t!]
    \centering    \includegraphics[width=\textwidth]{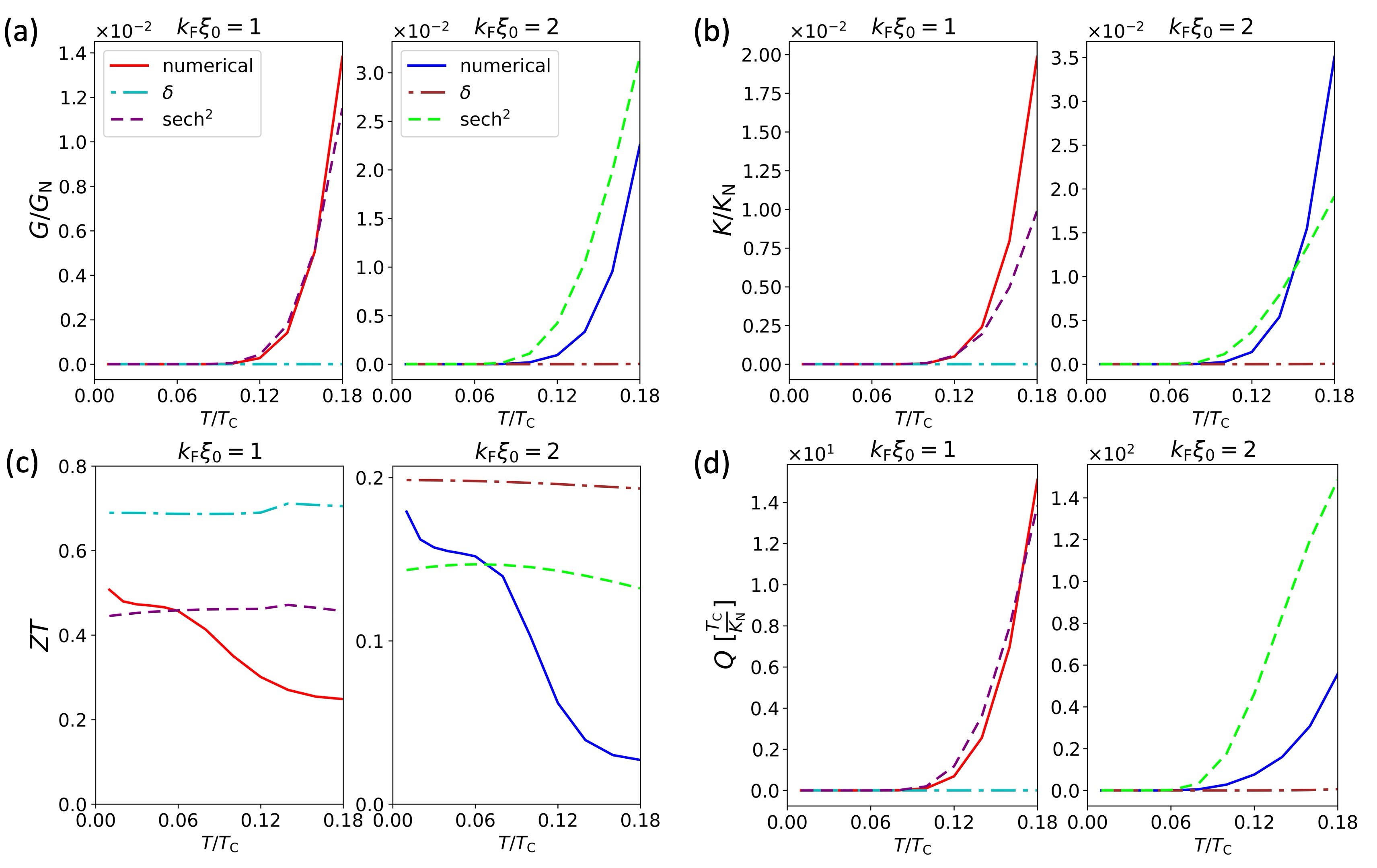}
    \caption{\textbf{SN Junction:} Electrical conductance \textbf{(a)}, thermal conductance \textbf{(b)}, figure of merit \textbf{(c)}, and power factor \textbf{(d)} are plotted versus normalized temperature for $k_{\rm F}\xi_0 = 1$ and $2$. Electrical (thermal) conductance is scaled by its corresponding value $G_{\rm N}$ ($K_{\rm N}$) for a Normal-Normal (NN) junction, with the same junction parameters, at the critical temperature. Power factor $Q$ is shown in the units of $T_{\rm C}/K_{\rm N}$. The numerical simulation result is plotted along with the delta function and $\sech^2$ approximation for the local density of states peaks.}
    \label{Fig2}
\end{figure*}

In the theory of thermoelectric response, a fundamental role is played by the local density of states (DOS),
\begin{equation}
    \rho_{\rm 2D}(E,\mathbf{r}) = \sum_n |u_n\r|^2 f'(E-E_n) + |v_n\r|^2f'(E+E_n),
    \label{ldos}
\end{equation}
where the subscript 2D refers to effective projection in two dimensions of the local DOS. We assume that we have integrated along the $z$-direction, which, in turn, changes the prefactor of the Onsager coefficients, as we will discuss in the next section. We omit the subscript 2D in the remainder of the article. As mentioned, a vortex makes $u_n$ and $v_n$ different for every energy eigenstate, leading to an asymmetric density of states structure (see Eq. \ref{ldos}). Our numerical estimate for the local density of states is shown in Fig \ref{Fig1} (c). The plot is in the subgap energy region; thus, the two peaks correspond to the bound states. The positive energy peak corresponds to the particle-like bound state, while the negative energy peak corresponds to the hole-like bound state. The asymmetry between the two peaks is evident, especially at $r=0$, where the particle-like peak starts with a high finite value and the hole-like peak begins with zero. The phase relation of $\Delta$ with $u$ and $v$ amplitudes ensures they are separated by one unit of angular momentum. Because of this difference, it turns out that the eigenstates of the BdG equation containing the zero angular momentum solution only have positive energy contributions in the local density of states, leading to this drastic asymmetry at the vortex core since only the zero angular momentum solution, $J_0(k_0 r)$, is non-zero at the origin. For more information, we refer the readers to the Supplementary Text. 
Thus, the vortex core is the region of the highest particle-hole asymmetry, which can be better exploited with an STM tip of Fig.~\ref{Fig1} (a).   

Furthermore, our results show that the source of this particle-hole asymmetry fundamentally lies in the asymmetry of the charge carriers present in the material. We assumed that the material has particle-like charge carriers; thus, we find symmetry breaking in favor of particles. If we initially assumed that the charge carriers are of hole type, we find that the asymmetry is reversed, and hole-like contributions to the bound states are more dominant.

In addition to providing more accurate numerical results, we also compare the numerical results with two simple semi-analytical approximations to calculate the thermoelectric coefficients. In both approximations, first of all, we assume that only a single VBS is present locally near the vortex. This leads to two different peaks in the local density of states, corresponding to the electron-like and hole-like contributions of the quasiparticle excitation (see Eq.~\ref{ldos}). In the first approximation, we further assume that these peaks are extremely sharp, represented by the Dirac-$\delta$ function. The benefit is that one can analytically calculate the temperature behavior of the thermoelectric coefficients up to a normalization factor. In the other approximation, we take the same peaks, but we assume a thermal broadening using $\sech^2((E_n \mp E)/(k_{\rm B}T))$, coming from the Fermi function derivative in the local density of states expression (Eq.~\ref{ldos}). In both models, the energies corresponding to VBS and the area under the VBS peaks are taken from the numerical solution to provide a meaningful comparison to the numerics. More details can be found in the Supplementary Text.

\vspace{1em}
\noindent\textbf{Thermoelectric Response}

\noindent We can now calculate the linear thermoelectric response of a single vortex. 
The electrical ($I$) and heat ($J$) current in the junction obey the Onsager relation,
\begin{equation}
\binom{I}{J}=\begin{pmatrix}
      L_{11}   &  L_{12} \\
      L_{21}  & L_{22} 
    \end{pmatrix}\binom{\Delta V/T}{\Delta T/T^2},
\end{equation}
where the voltage (temperature) difference  $\Delta V$ ($\Delta T$) across the junction determines the affinity $\Delta V/T$ ($\Delta T/T^2$)\cite{benenti2017fundamental}. The local Onsager matrix $L_{ij}$ is given by

\begin{equation}
    L_{ij}
    = -g T \int dE \int d^2{\bf r} |t({\bf r})|^2 \rho(E, {\bf r}) f'(E) \left(E/e\right)^{i+j-2},
    \label{onsager}
\end{equation}
where $i,j = 1,2$, and we assume that the density of states of the normal metal is flat over the relevant transport energy window and is thus incorporated into the prefactor $g$. The integration over the $z$-direction is absorbed into the prefactor $g$. The tunneling coupling inhomogeneities in the junction have been included in the energy-independent normalized tunneling prefactor $|t({\bf r})|^2$. For a junction covering the whole vortex $|t({\bf r})|^2=1$, whereas for an STM tip centered at ${\bf r_0}$, we model it as a Gaussian envelope $|t({\bf r})|^2=(2\pi R_{\rm STM}^2)^{-1}e^{|{\bf r}-{\bf r_0}|^2/2R_{\rm STM}^2}$, where $R_{\rm STM}$ is the effective STM tip radius. In the following, we mainly consider the former (SN junction) if it is not specified otherwise. 
Knowledge of $L_{ij}$ can be used to calculate all the different thermoelectric coefficients. 
For example, the Seebeck coefficient 
 (thermopower) is the voltage generated, divided by the temperature bias, across the junction given a temperature imbalance at zero total current,
\be
S = - \left.\frac{\Delta V}{\Delta T}\right\vert_{I =0} = \frac{1}{T} \frac{L_{12}}{L_{11}}.
\ee
Similarly, it is straightforward to find the electrical conductivity $G=L_{11}/T$, the open circuit thermal conductivity $K = (L_{22} - L_{21}L_{12}/L_{11})/T^2$ and other standard linear thermoelectric coefficients such as figure of merit $ZT = GS^2T/K$ and power factor $Q = GS^2$.

In Fig.~\ref{Fig1} (d), the left plot shows the numerical Seebeck coefficient (red solid line) plotted against temperature normalized with the critical temperature $T_C$ compared with the analytical results obtained with delta-like VBS (dot-dashed line) and effectively broadened VBS (dashed line). Significantly, we predict a sizeable thermoelectric response of a few ${\rm mV/K}$ for relatively small values of $k_{\rm F}\xi_0$. We show that increasing the parameter $k_{\rm F}\xi_0$ correspondingly reduces the thermoelectric response (see Fig.~\ref{Fig1} (d) right plot). Note also the sharp dependence of $S$ on the temperature in the low-temperature regime (much lower than the critical temperature). This feature can be exploited in vortex-localized temperature sensing at ultra-low temperatures.

Finally, it is also interesting to investigate the other aforementioned thermoelectric transport coefficients in Fig.~\ref{Fig2}, where we compare the numerical results with the two approximate analytical methods. In the first two panels, the electrical and thermal conductances show an exponential rise in temperature, as expected, since the bound states get exponentially more occupied. $ ZT$ decreases with the temperature increase, reaching a plateau around $T\approx 0.06\ T_C$, which may represent the optimal temperature value to measure the thermoelectric effect. However, a single vortex's power factor $Q$ increases exponentially with temperature since the exponential increase of $G$ with the temperature dominates the polynomial scaling of $S$. However, by increasing the number of vortices $N_v$ present in the junction, such as by slightly increasing the magnetic field, we expect that $G,K,Q$ scale with them proportionally while $S$ remains roughly the same. We can thus compensate for the low power factor by increasing the vortex density.

\begin{figure*}[t]
    \centering
    \includegraphics[width=\textwidth]{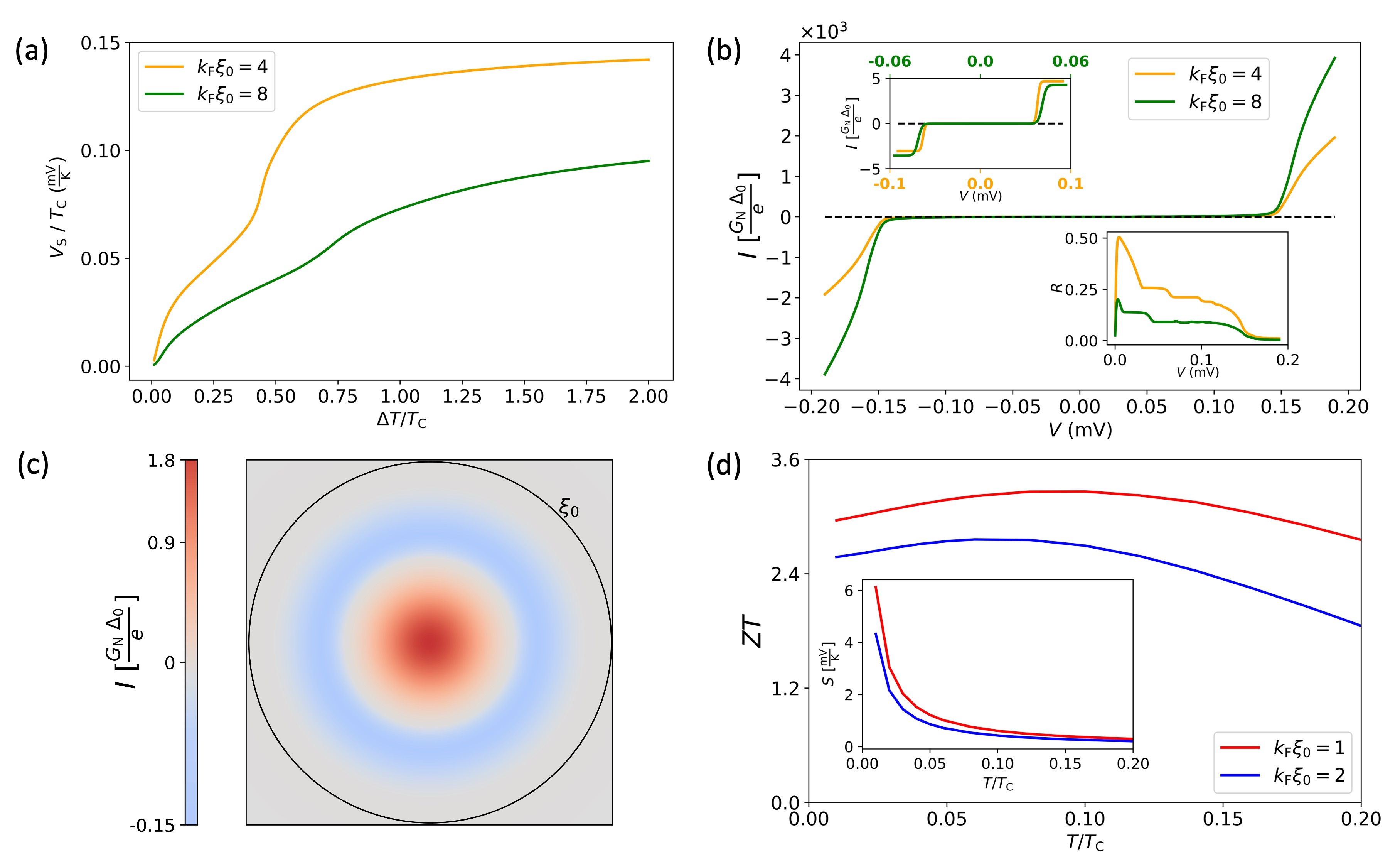}
    \caption{\textbf{(a)} Non-linear thermovoltage ($V_{\rm S}$) plotted w.r.t. the temperature bias across the NS junction. Two cases of $k_{\rm F} \xi_0$ ($ = 4,\ 8$) are shown, where the superconductor is kept at a fixed lower temperature of $0.05\ T_{\rm C}$. \textbf{(b)} I-V characteristic (no temperature bias) for the NS junction with $T = 0.01\ T_{\rm C}$ and $k_{\rm F} \xi_0 = 4,\ 8$. For the bigger range of $V\in [-0.2\ {\rm mV},0.2\ {\rm mV}]$, the plot looks reciprocal in current, but the upper inset plots illustrate the non-reciprocity in current at lower values of voltages (corresponding to the VBS). The lower inset shows the rectification factor plotted versus voltage for the I-V characteristic. \textbf{(c)} Electrical current going from different points in a square region surrounding the vortex (edge size $= 2\xi_0$) to a hotter STM tip (of effective radius $\xi_0/10$). The superconductor (STM tip) is kept at a temperature of $0.01\ T_{\rm C}$ ($0.21\ T_{\rm C}$). Current values in \textbf{(b,c)} are expressed in the units of $G_{\rm N}\Delta_0/e$, where $G_{\rm N}$ is the NN junction's electrical conductance at the critical temperature. \textbf{(d)} Numerical estimate for the figure of merit when the thermoelectric response is measured with an STM tip of size $\approx\xi_0/10$, for $k_{\rm F}\xi_0 = 1,\ 2$, plotted w.r.t. the temperature. The inset shows the variation of the Seebeck coefficient with temperature for the same parameters.}
    \label{Fig3}
\end{figure*}

We can calculate the non-linear thermoelectric response of a single vortex in the SN junction in the form of open circuit Seebeck thermovoltage $V_S$, which represents the voltage difference localized at the single vortex SN junction in an open circuit where the normal metal is heated at $ T+\Delta T$ and the superconductor is kept at base temperature $T$.
This is determined by solving
\begin{equation}
    \!\!\int\!\!dE \!\!\int \!\!d{\bf r}^2 \rho(E,{\bf r})(f_{\rm S}(E,T)
    - f_{\rm N}(E+eV_S,T+\Delta T)) = 0,
\end{equation}
which is not affected by the vortex number $N_v$. In Fig.~\ref{Fig3} (a), we show the evolution of the thermovoltage $V_S$ with the temperature difference $\Delta T$. 

Further, at thermal equilibrium, the electron-hole asymmetry in the vortex core expectedly also shows up in the breaking of the reciprocity ($I(V)=-I(-V)$ of the I-V characteristic for the single-vortex NS junction (Fig.~\ref{Fig3} (b)). 
Indeed, the current can be computed at different values of voltages by
\begin{equation}
    I = \frac{g}{e} \int \!\!dE\int\!\!d{\bf r}^2 \rho(E,{\bf r})(f_{\rm S}(E,T) - f_{\rm N}(E+eV,T)).
\end{equation}
Since we are mainly interested in the VBS contribution, we plot the IV characteristic around $V\approx0$. Looking at the general behavior, the IV characteristic still seems quite reciprocal because that is typical of the dominating supra gap high-energy contributions. However, zooming deeply into the subgap range (upper inset), we report a non-reciprocal behavior. This is better illustrated by the rectification factor ($R$, defined as $R~=~(I(V)-|I(-V)|)/(I(V)+|I(-V)|) \text{ for } V\ge0$, shown in the lower inset) versus voltage. Note that for the case of $k_{\rm F}\xi_0 = 4$, $R$ is greater than $0.2$ for a significant range of voltage in the subgap, peaking at $0.5$. Certainly, the weak non-reciprocity reported suggests that the thermoelectric features represent an optimal method to test the electron-hole asymmetry in the vortex core experimentally. This non-reciprocity can be used as an independent test for the reported asymmetry.

Before concluding, exploring how a local thermoelectric probe, such as an STM tip, would potentially affect the thermoelectric performances is interesting. We assume that the STM has an effective tunneling radius of $\xi_0/10$. A hot STM tip near the vortex leads to electrical current flow between the tip and the superconductor. In Fig.~\ref{Fig3} (c), this current is shown for a region surrounding the vortex - creating an {\it image} of the vortex. Near the vortex core, the current is positive (flowing from the tip to the superconductor) before becoming negative in a ring surrounding the vortex. The particle-dominant (hole-dominant) region has a positive (negative) current flow. A local probe at the vortex core is expected to be able to target the desired part of VBS without averaging over the hole-dominant regions, which emerge because of the oscillatory behavior of the Bessel functions. Indeed, we find that the Seebeck coefficient and ZT are higher for the local probe. However, the conductances are reduced (see Supplementary Text) because of decreased effective tunneling contact area. In Fig.~\ref{Fig3} (d), we have included the result for $ZT$ and $S$. $ZT$ with the STM setup significantly improves what we get with the SN junction. Further improvement of $ZT$ can be obtained with a smaller tip, sacrificing current and power. The supplementary material shows our results for the rest of the thermoelectric coefficients.

\vspace{1em}
\noindent\textbf{Conclusions}

\noindent Starting with the BdG equation to describe a single fluxon in a type-II superconductor, we saw how the nonlinear self-consistency requirement of the superconducting gap led to particle-hole symmetry breaking for bound states with energies within the superconducting gap.  The local density of states shows this particle-hole asymmetry is most significant at the vortex core. We proposed two setups to verify our prediction of a robust thermoelectric effect experimentally. Our results predict the Seebeck coefficient to be a few ${\rm mV/K}$ at sufficiently low $k_{\rm F}\xi_0$ and temperature for both experimental setups. Moreover, there is a sharp decline in $S$ with the increase in temperature, in contrast to the usual linear thermoelectric phenomena. The strong localized thermoelectricity could play an important role in highly localized bolometric applications or in building sensitive low-temperature thermocouples. The asymmetry is manifested not only in the thermovoltage but also in the non-reciprocal $I-V$ characteristic. The rectification factor above $0.2$ for $k_{\rm F}\xi_0 = 4$ makes the SN junction a potential candidate to build ultra-low temperature diodes. The existence of localized bound states can also be utilized in building sensitive, low-energy bolometers with short latency. In such cases, the radiation in the form of an incoming low-energy photon is absorbed by exciting quasiparticles collected in the vortex core, mainly affecting the bound states of the fluxon. The local hot-spot
could lead to the generation of thermovoltage spikes in a local junction.  An Abrikosov lattice of pinned vortices can then act as a multipixel camera, where each vortex acts as a pixel of the camera. Separate electrical contact to each vortex pixel allows \textit{passive} voltage detection of temperature spikes caused by multiple single-photon detection. 
\section*{Acknowledgments}
ANS, BB, and ANJ acknowledge the support of the U.S. Department of Energy (DOE), Office of Science, Basic Energy Sciences (BES), under Award No. DESC0017890. FG and AB acknowledge the EU’s Horizon 2020 Research and Innovation Framework Programme under Grant No. 964398 (SUPERGATE), No. 101057977 (SPECTRUM), and the PNRR MUR project PE0000023-NQSTI for partial financial support. AB
acknowledges MUR-PRIN 2022 - Grant No. 2022B9P8LN - (PE3)-Project NEThEQS  ``Non-equilibrium coherent thermal effects in quantum systems" in PNRR Mission 4 - Component 2 - Investment 1.1 ``Fondo per il Programma Nazionale di Ricerca e Progetti di Rilevante Interesse Nazionale (PRIN)" funded  by the European Union - Next Generation EU, the Royal Society through the International Exchanges between the UK and Italy (Grants No. IEC R2 192166) and CNR project QTHERMONANO.



    
\bibliography{naturemag}
\bibliographystyle{naturemag}

\clearpage
\section*{Supplementary Text}


\vspace{1em}

\noindent\textbf{Numerical Methods}

\noindent Our objective is to numerically solve the Bogoliubov-de-Gennes (BdG) equations, at low enough temperatures ($T/T_{\rm C} < 1/k_{\rm F}\xi_0)$, and magnetic field just slightly above the first critical field $H_{\rm c1}$, so that we have only a single vortex. For this weak magnetic field, we can ignore the dependence of vector-potential on the Hamiltonian\cite{de1999superconductivity}. Choosing the length scale to be coherence length $\xi_0$, the BdG equations can be written in the {\it dimensionless form}
\be
\begin{aligned}
&(\frac{-1}{2k_{\rm F}\xi_0}\nabla^2 - E_F)u_n(\mathbf{r}) + \Delta(\mathbf{r})v_n(\mathbf{r}) = E_nu_n(\mathbf{r}),\\
-&(\frac{-1}{2k_{\rm F}\xi_0}\nabla^2 - E_F)v_n(\mathbf{r}) + \Delta^*(\mathbf{r})u_n(\mathbf{r}) = E_nv_n(\mathbf{r}),
\end{aligned}
\label{BdG2}
\ee
where we have absorbed the mean field potential $U\r$ into the Fermi energy $E_{\rm F}$. Applying the gauge $\Delta\r=|\Delta\r|e^{-i\theta}$, we can write the quasiparticle amplitudes in terms of angular momentum eigenstates,
\be
\begin{aligned}
u_n(r,\theta) &= \frac{u_{nm}(r)}{\sqrt{2\pi}}e^{im\theta},\\
v_n(r,\theta) &= \frac{v_{nm}(r)}{\sqrt{2\pi}}e^{i(m+1)\theta}.
\label{uv}
\end{aligned}
\ee
The index $n$ denotes the eigenstate labelling and $m$ is connected to the angular momentum $\mu = m+1/2$ of the solution.  

Bardeen et al. \cite{bardeen1969structure} provided a semi-analytic model that we apply to this problem by approximating the pair potential as zero in the vortex core and making use of the Fourier-Bessel expansion to express the quasiparticle amplitudes in an orthonormal basis. To arrive at more accurate results, we solve the problem numerically using self-consistency calculations\cite{gygi1991selfconsistent,hayashi1998low,kato2000quasi,edblom2012numerical,hu2011local}. We start with a constant pair potential as our ansatz, and calculate the radial part of the quasiparticle amplitudes $u_{nm}(r)$, $v_{nm}(r)$ using equations~(\ref{BdG2}) and~(\ref{uv}). They are, in turn, used to calculate the pair potential, and this cycle is repeated until we achieve the convergence up to some tolerance.

Exploiting the cylindrical symmetry, the radial parts can be expressed in terms of Fourier-Bessel series expansion\cite{kato2000quasi,gygi1991selfconsistent,hu2011local,governale2020finite},
\be
\begin{aligned}
    &u_{nm}(r) = \sum_j c_{nj}\phi_{jm}(r),\\
    &v_{nm}(r) = \sum_j d_{nj}\phi_{j(m+1)}(r),
    \label{uvphi}
\end{aligned}
\ee
where $\phi_{jm}$'s are Bessel functions normalized over a disc of radius $R$,
\be
\phi_{j m}(r)=\frac{\sqrt{2}}{R J_{m+1}\left(\alpha_{j m}\right)} J_m\left(\frac{\alpha_{j m}}{R} r\right),
\ee
and $\alpha_{jm}$ is the $j^{\rm th}$ zero of $J_m$. Note that the basis functions $u_{nm},v_{nm}$ are zero at the boundary of the cylinder. Importantly, the particle-like lowest energy solution $u_{1/2}$ behaves as $J_0(k r)$, which limits to 1 as $r=0$, while the hole-like solution $v_{1/2}$ behaves as $J_1(k r)$, which limits to 0 as $r=0$.  Conversely, the $\mu=-1/2$ solution has a non-zero $v_{-1/2}$ solution at $r=0$, but the $u_{-1/2}$ solution is zero at the origin.  However, the negative angular momentum has negative energy, so the role of $u, v$ switch  - both the $u_{1/2}$ and $v_{-1/2}$ solution have positive energy and are non-zero at the core of the vortex. In contrast, all other solutions vanish, making the vortex core dominated by particle-like components. The basis functions $\phi_{jm}$'s form a complete orthonormal set on $r\in[0,R]$,
\be
\int_0^Rdr\ r \phi_{jm}(r) \phi_{j'm}(r) = \delta_{jj'}.
\ee
Quasiparticle amplitudes expansion in Eq. \ref{uvphi} can be used to simplify the BdG equation \cite{edblom2012numerical} to give a matrix eigenvalue equation of the form,
\be
\begin{pmatrix}
    T^m & D \\ D^{\rm T} & -T^{m+1}
\end{pmatrix} \Psi_n = E_n\Psi_n,
\label{eigen}
\ee
where $T^{m}$ corresponds to the kinetic energy term in the Hamiltonian and is diagonal with elements,
\begin{equation}
    T_{j j}^{m}=\frac{1}{2 k_{\rm F} \xi_0}\left(\frac{\alpha_{j m}}{R}\right)^2-E_F.
\end{equation}
$D$ contains the coupling due to the pair potential and is given by,
\begin{equation}
    D_{j^{\prime} j}=\int_0^R \mathrm{~d} r\ r \phi_{j^{\prime} m}(r) \Delta(r) \phi_{j (m+1)}.
\end{equation}
Using Eq. \ref{uv} and \ref{uvphi}, we can write the order parameter as
\begin{multline}
\Delta (r) = V \sum_{\substack{m \\ E_{nm}\leq E_c}}\sum_{j_1j_2}c_{nj_1}d_{nj_2}\\
\phi_{j_1m}(r)\phi_{j_2(m+1)}(r)\left(1-2f\left(E_{nm}\right)\right).
\end{multline}
We impose a cut-off on the magnitude of energy while doing a summation over the energy eigenstates to save computation time. The cut-off is taken to be high enough not to affect the calculated results. To be specific, we have chosen the cutoff energy $E_{\rm C}$ to be $5k_{\rm B}T_{\rm C}$ for all the different cases of $k_{\rm F}\xi_0$ in our analysis.

For a detailed description of the numerical algorithm used to solve this problem, we refer the readers to Ref.~[\onlinecite{edblom2012numerical}], but we mention here the essential steps involved. To save on computation time, all the functions of $r$ are written in terms of lists of $1000$ equidistant points in space. We ensured that the results converged by comparing them with the $5000$ points case and finding that the differences were negligible. We start with a guess for the pair potential as the constant BCS value ($=1.764k_{\rm B}T_{\rm C})$. Using the eigenvalue equation~(\ref{eigen}), $c_{nj}$'s and $d_{nj}$'s are found out, which are then used to calculate the next guess for $\Delta(r)$. This cycle is repeated until the convergence is reached. In general, this converged result does not conserve particle number (when compared with the normal metal value), where the particle number is calculated by
\begin{equation}
    N = 2 \sum_{nj}[|c_{nj}|^2f_n + |d_{nj}|^2(1-f_n)].
\end{equation}
After each convergence, $E_{\rm F}$ is slightly varied to nudge the particle number closer to the desired value. When done multiple times, we obtain the value of $E_{\rm F}$ (or chemical potential), which conserves the particle number. This value is different for different temperatures, and this process is thus repeated for each temperature value.

Integrals over energy and space are required to calculate the thermoelectric coefficients, thermovoltage, or current. Numerical integration is approximate, but we ensured that the relative error at each data point is less than $10^{-3}$ by taking enough summation points. The numerics are also approximate in calculating the density of states for energies higher than the band gap, especially for lower values of $k_{\rm F}\xi_0$. Since those states should not contribute to the thermoelectric coefficients (because of particle-hole symmetry outside of the band gap), we only consider the bound states while calculating the Onsager coefficients. While calculating current and thermovoltage, we consider higher values of $k_{\rm F}\xi_0$ and thus, the errors are not expected to be significant for the results shown in the article. 
\vspace{1em}

\noindent\textbf{Analytical Models}

\noindent In Fig.~\ref{Fig1} (d) and Fig.~\ref{Fig2}, along with the numerical results, some semi-analytical results are also plotted based on two different approximations - $\delta$ and $\sech^2$. The former consists of the assumption that the local density of states is a combination of different Dirac-delta functions centered at bound state energy values, weighted by the quasiparticle amplitudes,
\begin{equation}
    \rho(E, {\bf r}) =\sum_n |u_n({\bf r})|^2 \delta(E - E_n) + |v_n({\bf r})|^2 \delta(E+E_n).
\end{equation}
We further restrict this model to consider only the lowest energy bound states, corresponding to $\mu=1/2$, and thus giving,
\begin{equation}
    \rho(E) = \int d^2\mathbf{r} \rho(E,\mathbf{r}) = \lambda_+ \delta(E-E_{1/2}) + \lambda_- \delta(E+E_{1/2}),
\end{equation}
where $\lambda_+ = \sum_j |c_{1/2,j}|^2$ and $\lambda_- = \sum_j |d_{1/2,j}|^2$. Plugging this expression into equation~(\ref{onsager}), we get the Onsager matrix as,
\begin{equation}
    \begin{aligned}
    L = &gT(-f^\prime(E_{1/2}))\cross\\
    &\begin{pmatrix}
      \lambda_+ + \lambda_-  &  (\lambda_+ - \lambda_-)E_{1/2}/e \\
      (\lambda_+ - \lambda_-)E_{1/2}/e & (\lambda_+ + \lambda_-)(E_{1/2}/e)^2 
    \end{pmatrix}    
    \end{aligned}
\end{equation}
\vspace{0.5em}

The thermoelectric coefficients then have the form,
\begin{equation}
    S = \frac{E_{1/2}}{e T} \frac{\lambda_+ - \lambda_-}{\lambda_+ + \lambda_-},
\end{equation}
\begin{equation}
    G = g (\lambda_+ + \lambda_-) (-f'(E_{1/2})),
\end{equation}

\begin{equation}
    K = \frac{g  (-f'(E_{1/2})) E_{1/2}^2}{T e^2} \frac{ 4 \lambda_+ \lambda_-}{\lambda_+ +\lambda_+ },
\end{equation}
\begin{equation}
    ZT = \frac{(\lambda_+ - \lambda_-)^2}{4\lambda_+\lambda_-},
\end{equation}
\begin{equation}
\begin{aligned}
    \text{and } Q =\  g \frac{E_{1/2}^2}{e^2T^2}(\lambda_+ + \lambda_-)(\lambda_+ - \lambda_-)^2 (-f'(E_{1/2})).
\end{aligned}
\end{equation}
Note that this approximate density of states is unit normalized, keeping in account only the lowest energy state contribution and neglecting any higher energy states contributions. Thus, a normalization factor needs to be explicitly fixed to compare with the numerical results. This can be understood as one of the main limitation of the semi-analytical models.

The other approximated low-energy model assumes that the delta-like peaks in the density of states are spread due to a sort of thermal broadening which takes the form $\sech^2(x)$ function peaked in energy, again centered on the lowest energy-bound state previous considered, such as
\begin{equation}
\begin{aligned}
    \rho(E) = \ &\frac{\lambda_+}{4k_BT} \sech^2\left(\frac{E-E_{1/2}}{2k_BT}\right)\\
    &+ \frac{\lambda_-}{4k_BT} \sech^2\left(\frac{E+E_{1/2}}{2k_BT}\right).
\end{aligned}
\end{equation}
The inspiration to consider $\sech^2$ indeed comes from equation~(\ref{ldos}), where the derivative of the Fermi function results in $\sech^2$ thermal broadened peaks. Again, the density of states is unit-normalized, so this model also needs to be fixed by properly matching the normalization factor with the numerical results.

\vspace{1em}

\noindent\textbf{STM Junction: Thermoelectric Coefficients}

\begin{figure}[h]
    \centering
    \includegraphics[width=0.5\textwidth]{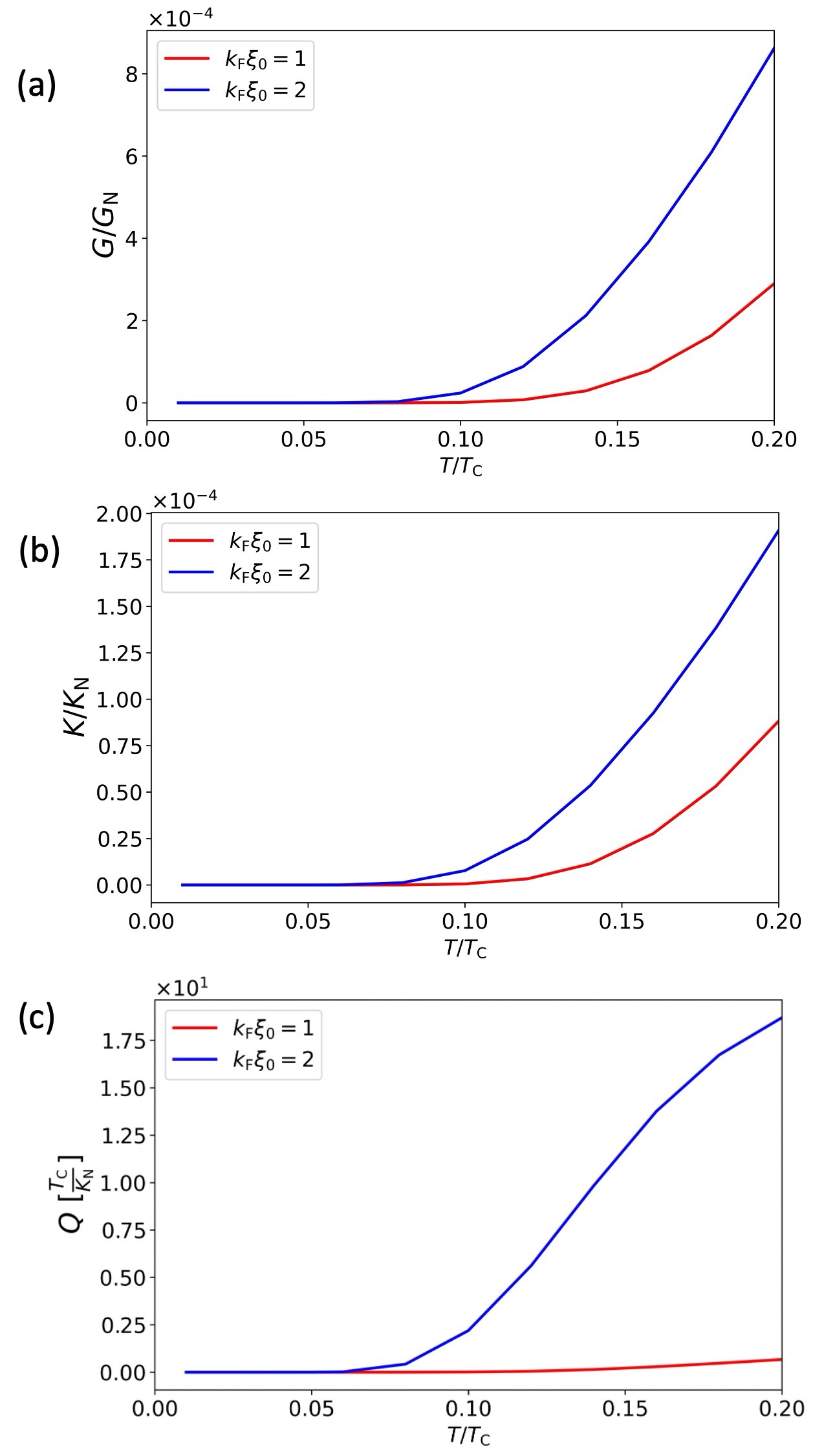}
    \caption{\textbf{STM Junction:} Electrical conductance \textbf{(a)}, thermal conductance \textbf{(b)}, and power factor \textbf{(c)} are plotted versus normalized temperature for $k_{\rm F}\xi_0 = 1$ and $k_{\rm F}\xi_0 = 2$, scaled by the factor $G_{\rm N}$, $K_{\rm N}$, and $K_{\rm N}/T_{\rm C}$ respectively.}
    \label{Supp_Fig}
\end{figure}

\noindent In the main text, we only showed the Seebeck coefficient and the figure of merit for the STM junction setup. In Fig.~\ref{Fig4}, we have also included the other thermoelectric transport coefficients ($G$, $K$, and $Q$). We see that $G$, $K$, and $Q$ behave quite similarly to their counterpart in the SN junction, but they have much lower absolute values, owing to the decreased effective area of the thermoelectric junction.

\end{document}